\documentclass[a4paper,11pt]{article}
\pdfoutput=1 

\usepackage{jcappub} 

\usepackage[T1]{fontenc} 
\usepackage{graphicx}
\usepackage{dcolumn}
\usepackage{bm}
\usepackage{amsmath,multirow}
\usepackage{amssymb}
\usepackage{epstopdf,tabularx}

\title{\boldmath Efficient high-order explicit symplectic splitting methods
	for post-Newtonian Hamiltonian systems}

\author[a,b]{Yujie Liang}
\author[a,b,1]{and Lijie Mei\note{Corresponding author.}}

\affiliation[a]{School of Mathematics, Yunnan Normal
	University, Kunming 650500, China}
\affiliation[b]{Yunnan Key Laboratory of Modern Analytical Mathematics
	and Applications, Yunnan Normal University, Kunming 650500, China}

\emailAdd{936010207@qq.com}
\emailAdd{bxhanm@126.com}

\abstract{
	The nonseparability of post-Newtonian (PN) Hamiltonian systems typically
necessitates the use of computationally expensive implicit integrators.
Recent research overcomes this limitation by embedding the dynamics
into a doubled phase space, which enables the development of
explicit symplectic methods.
However, existing specially designed explicit integrators suffer from
order reduction for high-order methods when the time stepsize is small,
i.e., $h <\varepsilon^3$. In this paper, we propose a novel extension
and splitting approach for the doubled Hamiltonian, under which
specially designed explicit symplectic integrators can be constructed.
It is shown that the proposed integrators achieve genuine high-order
convergence without order reduction and take advantage of the small
PN parameter $\varepsilon$. Numerical results from simulations
with 2PN spinning binaries demonstrate superior long-term conservation
of invariants and significantly higher computational efficiency compared
to both implicit methods and existing explicit splitting techniques.}

\begin{document}
	\maketitle
	\flushbottom
	
	\section{Introduction}
	
The post-Newtonian (PN) Hamiltonian approach provides a widely adopted
framework for modeling compact object systems consisting of neutron stars
and black holes to approximate general relativity in the slow-motion,
weak-field limit \cite{Damour2001,Andrade2001,Levi2014,Wu2015,Huang2014,Dubeibe2017,HuangLi2018}.
Since analytical methods are not applicable to most PN
Hamiltonian systems, numerical methods, especially symplectic methods that
guarantee good long-term behavior, are promising tools for studying the
dynamics of PN Hamiltonian systems \cite{Feng2010,Gauckler2018,Hairer2006}.

However, due to the nonseparability of PN Hamiltonian systems, classical
symplectic methods --- mainly including Gauss-type symplectic Runge--Kutta
methods \cite{Lasagni1988,SanzSerna1988,Suris1988}, mixed symplectic methods \cite{Lubich2010,Zhong2010,Mei2013}, symplectic generalized flow-composed
methods \cite{Antonana2022,Huang2024}, and semiexplicit symplectic methods \cite{Jayawardana2023,Ohsawa2023} --- are all implicit
\cite{McLachlan1992,McLachlan1993,Yoshida1993} and thus
suffer from low computational efficiency when long-term numerical
simulations are required. That is, despite several explicit symplectic
methods for some specific types of nonseparable Hamiltonian systems
\cite{Chin2009,Blanes2002,Tao2016b,McLachlan2004}, traditional
approaches cannot construct explicit symplectic methods for
general nonseparable Hamiltonian systems.

To efficiently integrate the PN Hamiltonian system, several attempts have
been made to produce explicit methods possessing good long-term behavior.
For example, the explicit methods in the doubled phase space framework
\cite{Liu2016,Luo2017a,Luo2017b,Luo2021,Pan2021} (based on the approach
proposed by Pihajoki \cite{Pihajoki2015}) and the near-symplectic methods
\cite{Mei2024}. Recently, Mei et al. \cite{Mei2025} have proved the
symplecticity and good long-term behavior of the explicit projection methods
for general nonseparable Hamiltonian systems. The authors of \cite{Niu2026}
further proposed some specially designed explicit methods that take into
account the small PN parameter $\varepsilon$, so that the global error is
much smaller than that of the originally proposed methods of the same order
in \cite{Mei2025}.

However, as noted in \cite{Niu2026}, the method EH4 established therein
suffers from order reduction when $h<\varepsilon^3$. That is, it is of
order 4 for the case $h\geq\varepsilon^3$, while it is only of order
three when $h<\varepsilon^3$. Consequently, all explicit methods of
order higher than 4 established following the idea in \cite{Niu2026}
share the same order reduction issue as EH4. Hence, the construction of
high-order, specially designed, efficient explicit methods without order
reduction under any condition still remains a challenge for PN
Hamiltonian systems.

In this paper, we first consider the PN Hamiltonian in the doubled phase
space, where the Hamiltonian is extended not by a simple doubled copy
as in \cite{Pihajoki2015}, but in a specially designed manner.
Then, we present a three-operator Hamiltonian splitting approach and
produce second-order and fourth-order explicit symplectic splitting methods
for the doubled Hamiltonian. By adding a special explicit symplectic
projection to the explicit splitting methods, we finally obtain explicit
methods that are both symplectic and genuine fourth-order for the PN
Hamiltonian system. Following this idea, higher-order explicit methods can
be similarly obtained just by using a higher-order Hamiltonian splitting
for the doubled Hamiltonian.

This paper is organized as follows. In Section \ref{sec:explicit_methods},
we review the explicit symplectic method with special splitting originally
proposed in \cite{Niu2026}. Inspired by this approach, we introduce a novel
explicit symplectic integrator with a special three-operator Hamiltonian
splitting. Section \ref{sec:formula} presents the 2PN Hamiltonian
model for a compact binary system with spin effects. In Section \ref{sec:numerical}
we conduct numerical experiments to demonstrate the superiority of the proposed
explicit symplectic algorithm in terms of convergence, computational efficiency,
and long-term energy behavior. Finally, Section \ref{sec:conclusion} summarizes
the contributions of this paper.
	
	\section{Explicit symplectic integrators}\label{sec:explicit_methods}
In this section, we first review the pseudo-high-order symplectic
method proposed in \cite{Niu2026} and discuss its order
reduction issue. Then, we present a novel operator-splitting
approach that decomposes the doubled Hamiltonian into three
sub-Hamiltonians and derive the associated explicit methods.
We further show that the proposed methods do not encounter
the order reduction issue.

Since the PN Hamiltonian possesses a particular structure consisting
of a completely integrable part and a small PN correction part, we
conveniently write it in the following abstract form:
\begin{equation}\label{perturbatedHamilton}
	H(\bm{z})=H_0(\bm{z}) + \varepsilon H_1(\bm{z}),
\end{equation}
with the initial condition $\bm{z}(0)=\bm{z}_0$. Here, $\bm{z}=(\bm{p},
\bm{q})\in \mathbb{R}^{2d}$,
where $\bm{p}\in\mathbb{R}^{d}$ and $\bm{q}\in\mathbb{R}^{d}$
represent the generalized momenta
and positions, respectively, which are canonically conjugate to
each other. In general, $H_{0}$ is the Hamiltonian of the integrable
two-body Kepler problem, $H_{1}$ incorporates velocity- and
position-dependent PN/relativistic corrections, and the PN
parameter $\varepsilon \sim 1/c^2$ characterizes the scale of the PN effect.

\subsection{Pseudo-high-order explicit symplectic methods}
The methodology proposed in \cite{Niu2026} is now presented as follows.
Suppose that $h$ is the time stepsize. Since $H_0$ is completely integrable,
it is solvable in theory and we denote its phase flow by $\phi_t$.
Next, we consider an explicit symplectic solver for the nonintegrable and
nonseparable part  $H_1$.

Due to the nonseparability of $H_1$, we follow the idea described in
\cite{Mei2025} to construct symplectic solvers for it. By introducing
an auxiliary pair of conjugate variables $(\bm{x},\bm{y})\in\mathbb{R}^{2d}$,
the Hamiltonian $H_1(\bm{p},\bm{q})$ is extended to
\begin{equation}\label{eq:split_HPN}
	\widetilde{H}(\bm{p}, \bm{q}, \bm{x}, \bm{y}) = H_1(\bm{p}, \bm{y})
	+ H_1(\bm{x}, \bm{q}),
\end{equation}
whose canonical equations are
\begin{equation}\label{eq:extended_dynamics}
	\left\{
	\begin{aligned}
		\frac{\mathrm{d}\bm{p}}{\mathrm{d}t} &= f(\bm{x}, \bm{q}),
		&\quad \frac{\mathrm{d}\bm{q}}{\mathrm{d}t} &= g(\bm{p}, \bm{y}),
		\\
		\frac{\mathrm{d}\bm{x}}{\mathrm{d}t} &= f(\bm{p}, \bm{y}),
		&\quad \frac{\mathrm{d}\bm{y}}{\mathrm{d}t} &= g(\bm{x}, \bm{q}),
	\end{aligned}
	\right.
\end{equation}
where $f = -\frac{\partial H_1}{\partial \bm{q}}$ and
$g = +\frac{\partial H_1}{\partial \bm{p}}$.

It is obvious that the extended Hamiltonian $\widetilde{H}$
defined in \eqref{eq:split_HPN} is separable. By setting
$H_A=H_1(\bm{p}, \bm{y})$ and $H_B=H_1(\bm{x}, \bm{q})$,
the phase flows, which are respectively denoted by $\exp(t X_A)$
and $\exp(t X_B)$, admit the following closed-form solutions:
\begin{equation}\label{subflows}
	\begin{aligned}
		\exp(t X_{A}) &: (\bm{p}_0, \bm{q}_0,\bm{x}_0, \bm{y}_0)
		\mapsto
		&\big(
		\bm{p}_0,\;
		\bm{q}_0 + t\,g(\bm{p}_0, \bm{y}_0),\;
		\bm{x}_0 + t\,f(\bm{p}_0, \bm{y}_0),\;	
		\bm{y}_0
		\big), \\
		\exp(t X_{B}) &: (\bm{p}_0, \bm{q}_0, \bm{x}_0, \bm{y}_0)
		\mapsto
		&\big(
		\bm{p}_0 + t\,f(\bm{x}_0, \bm{q}_0),\;	
		\bm{q}_0 ,\;
		\bm{x}_0,\;
		\bm{y}_0  + t\,g(\bm{x}_0, \bm{q}_0)
		\big),
	\end{aligned}
\end{equation}
where $(\bm{p}_0, \bm{q}_0, \bm{x}_0, \bm{y}_0)$ are the
initial values, and $X_A$ and $X_B$ are Hamiltonian vector fields
defined via Poisson brackets:
\begin{equation*}
	\begin{aligned}
		X_{A}&:=\{\,\cdot\,,H_A\}=\frac{\partial H_A}{\partial \bm{p}}\frac{\partial }{\partial \bm{q}}
		-\frac{\partial H_A}{\partial \bm{y}}\frac{\partial }{\partial \bm{x}},
		\\
		X_{B}&:=\{\,\cdot\,,H_B\}=\frac{\partial H_B}{\partial \bm{x}}\frac{\partial }{\partial \bm{y}}
		-\frac{\partial H_B}{\partial \bm{q}}\frac{\partial }{\partial \bm{p}}.
	\end{aligned}
\end{equation*}
Then, a second-order explicit symplectic splitting method for the
extended Hamiltonian $\widetilde{H}(\bm{p}, \bm{q}, \bm{x}, \bm{y})$
with stepsize $h$ can be formulated as
\begin{equation}\label{extended_splitting}
	\text{EPN2}(h)=\exp\big(\tfrac{h}{2}X_B\big)\circ\exp(hX_A)\circ
	\exp\big(\tfrac{h}{2}X_B\big).
\end{equation}
Throughout this paper, we emphasize that \textit{the operators in the numerical method are composed in a right-to-left order}.

Following \cite{Mei2025}, we introduce the specific submanifold
$\mathcal{N} \subset \mathbb{R}^{4d}$ defined as:
\begin{equation*}
	\mathcal{N} = \bigl\{ (\bm{p}, \bm{q},\bm{x}, \bm{y})
	\in \mathbb{R}^{4d}\mid \bm{x} = \bm{p},\; \bm{y} = \bm{q} \bigr\}.
\end{equation*}
To map the solution in the doubled phase space $\mathbb{R}^{4d}$ to the
original phase space $\mathbb{R}^{2d}$, we further introduce the projection
mapping $\mathcal{P}: \mathbb{R}^{4d}\to
\mathcal{N}$, which is defined as follows:
\begin{equation}\label{eq:ED2_projection_row}
	(\widetilde{\bm{p}}, \widetilde{\bm{q}}) =
	\begin{cases}
		\bigl( \lambda_0 \bm{p} + (1-\lambda_0)\bm{x}, \;
		\mu_0 \bm{q} + (1-\mu_0)\bm{y}\bigr),
		& \text{if } n \text{ is odd}, \\[1.2ex]
		\bigl( \mu_0 \bm{p} + (1-\mu_0)\bm{x}, \;
		\lambda_0 \bm{q} + (1-\lambda_0)\bm{y}\bigr),
		& \text{if } n \text{ is even},
	\end{cases}
\end{equation}
where $(\widetilde{\bm{p}},\widetilde{\bm{q}},\widetilde{\bm{p}},
\widetilde{\bm{q}})=\mathcal{P}(\bm{p}, \bm{q},\bm{x}, \bm{y})$,
and $\lambda_0$ and $\mu_0$ are two free parameters whose values are
usually taken from the interval $(0,1)$.

Incorporating the projection operator $\mathcal{P}$ into the
explicit splitting method \eqref{extended_splitting} finally
yields the explicit integrator for the PN Hamiltonian
$\varepsilon H_1(\bm{p}, \bm{q})$ defined in
\eqref{perturbatedHamilton} as follows:
\begin{equation*}
	\begin{split}
		\text{ED2}_{\mathrm{PN}}(h) = \mathcal{P}\circ\exp\big(\tfrac{\varepsilon h}{2}X_B\big)
		\circ\exp(\varepsilon h X_A) \circ \exp\big(\tfrac{\varepsilon h}{2}X_B\big).
	\end{split}
\end{equation*}
Then, the St\"{o}rmer--Verlet (or leap-frog) scheme involving
the phase flow $\phi_h$ and the explicit solver $\text{ED2}_{\mathrm{PN}}$
yields the explicit integrator
\begin{equation}\label{ES2}
	\text{ES2}(h)=\phi_{h/2}\circ \text{ED2}_{\mathrm{PN}}(h)\circ\phi_{h/2},
\end{equation}
which is just the same as that defined in \cite{Niu2026}. Further
application of Yoshida's triple symmetric composition \cite{Yoshida1990}
and merging of neighbouring $\phi_t$ flows for $\text{ES2}$ yields
$\text{ES4}$:
\begin{eqnarray}
	\text{ES4}(h) &=& \phi_{\lambda h/2} \circ \mathrm{ED2}_{\mathrm{PN}}(\lambda h)
	\circ \phi_{(1-\lambda) h/2}\circ \mathrm{ED2}_{\mathrm{PN}}\big((1-2\lambda)h\big) \circ
	\phi_{(1-\lambda) h/2}
	\nonumber
	\\
	&&\circ \mathrm{ED2}_{\mathrm{PN}}(\lambda h)
	\circ \phi_{\lambda h/2},
	\label{eq:ES4}
\end{eqnarray}
where $\lambda = 1/(2 - 2^{1/3})$.
Here, we do not present their detailed numerical schemes, which can be found in \cite{Niu2026}.

As shown in \cite{Mei2025,Niu2026}, the explicit methods
$\text{ES2}$ and $\text{ES4}$ are symplectic and exhibit
excellent long-term energy conservation and linear global
error growth. However, although $\text{ES2}$ is truly
second-order in any case, $\text{ES4}$ suffers from order
reduction. The details are as follows.

Let $X_{0}=\{\,\cdot\,,H_0\}=\frac{\partial H_0}{\partial \bm{p}}
\frac{\partial }{\partial \bm{q}}-\frac{\partial H_0}{\partial \bm{q}}
\frac{\partial }{\partial \bm{p}}$,
$X_{1}=\{\,\cdot\,,H_1\}=\frac{\partial H_1}{\partial \bm{p}}
\frac{\partial }{\partial \bm{q}}-\frac{\partial H_1}{\partial \bm{q}}
\frac{\partial }{\partial \bm{p}}$, and $X=X_0+\varepsilon X_1$.
Then the phase flows of $H_0$, $H_1$, and $H$ can be
abstractly expressed by the exponential mappings
$\phi_t=\exp(tX_0)$, $\exp(tX_1)$, and $\exp(tX)$. We also suppose
that $H_0$ and $H_1$ have the same magnitude regardless of
the PN parameter $\varepsilon$ in the latter.

If an $r$th-order method with stepsize $h$ is applied to the Hamiltonian $H_1(z)$, then the local truncation
error in a single step is of order $\mathcal{O}(h^{r+1})$. Here we present a
more rigorous analysis to show that when the Hamiltonian being
numerically integrated is scaled by a factor $\varepsilon$, i.e., $\varepsilon H_1(z)$ rather than $H_1(z)$,
the local truncation error becomes
$\mathcal{O}(\varepsilon^{r+1} h^{r+1})$.

For the Hamiltonian $\varepsilon H_1(\bm{z})$,
the canonical equations are
\begin{equation}\label{err-analy-eq1}
	\frac{\mathrm{d}\bm{z}}{\mathrm{d}t}
	= \varepsilon J^{-1} \nabla H_1(\bm{z}).
\end{equation}
If we apply the time transformation $\tau = \varepsilon t$ to \eqref{err-analy-eq1}, the equations become
\begin{equation}\label{err-analy-eq2}
	\frac{\mathrm{d}\bm{z}}{\mathrm{d}\tau} = J^{-1} \nabla H_1(\bm{z}),
\end{equation}
which has the same form as that for the Hamiltonian $H_1(\bm{z})$, except
that the time variable becomes $\tau$ rather than $t$.
That is, applying an $r$th-order
method with stepsize $h$ to the system \eqref{err-analy-eq1} is equivalent to
applying the same method but with stepsize $\tilde{h} = \varepsilon h$ to the system \eqref{err-analy-eq2}.
Therefore, the local truncation error will be $\mathcal{O}(\tilde{h}^{r+1}) = \mathcal{O}(\varepsilon^{r+1} h^{r+1})$ when
applying an $r$th-order method to numerically solve the Hamiltonian $\varepsilon H_1(\bm{z})$,
and the $r$th-order method can be formulated by an exponential mapping as follows:
\begin{equation*}
	\exp\big( \varepsilon h X_1 + \mathcal{O}(\varepsilon^{r+1} h^{r+1}) \big).
\end{equation*}

According to the results on projection methods \cite{Hairer2006}, the explicit
projection operator defined by \eqref{eq:ED2_projection_row} does not
change the accuracy of a numerical method. Based on the above analysis,
we conclude that the explicit solver $\text{ED2}_{\mathrm{PN}}$ for $\varepsilon H_1(\bm{z})$ satisfies
\begin{equation*}
	\text{ED2}_{\mathrm{PN}}(h)=\exp\big(\varepsilon h X_1 + \varepsilon^{3} h^{3} \widetilde{D} + \mathcal{O}(\varepsilon^{4} h^{4})\big),
\end{equation*}
where the vector field $\widetilde{D}$ is determined by $X_A$ and $X_B$.

Then, it follows from the BCH formula (see in Chap. III.4 on pp. 83 of \cite{Hairer2006})
that the St\"{o}rmer--Verlet scheme
ES2 defined by \eqref{ES2} possesses the following expression:
\begin{eqnarray}
	\text{ES2}(h)=&&\exp(\tfrac{h}{2}X_0)\circ \exp\big(\varepsilon h X_1 +
	\mathcal{O}(\varepsilon^{3}h^{3})\big)\circ\exp(\tfrac{h}{2}X_0)
	\nonumber
	\\
	=&&\exp\big(hX + \varepsilon h^{3} \widetilde{S} + \varepsilon^{3} h^{3} \widetilde{D}+ \mathcal{O}(\varepsilon^{4} h^{4})
	+ \mathcal{O}(\varepsilon h^{5})
	+\cdots \big),
	\label{ES2-error}
\end{eqnarray}
where $\widetilde{S}=-\frac{1}{24}[X_0,[X_0,X_1]] + \frac{1}{12}\varepsilon[X_1,[X_1,X_0]]$ and
$[\cdot,\cdot]$ denotes the commutator defined by $[A,B]=AB-BA$.
For the method ES4 defined by \eqref{eq:ES4}, the terms $\varepsilon h^{3} \widetilde{S}$
and $\varepsilon^{3} h^{3} \widetilde{D}$ vanish because $2\lambda^3+(1-2\lambda)^3=0$ holds for
$\lambda={1}/{(2-2^{1/3})}$. However, the term
$\mathcal{O}(\varepsilon^{4} h^{4})$ does not vanish due to the fact that
$2\lambda^4+(1-2\lambda)^4\neq0$. This leads to the expression:
\begin{equation}\label{ES4-error}
	\text{ES4}(h)=\exp\big(hX
	+\mathcal{O}(\varepsilon^{4} h^{4})
	+\mathcal{O}(\varepsilon h^{5})+\cdots \big).
\end{equation}

The estimate in \eqref{ES2-error} clearly indicates the second-order accuracy of
ES2 regardless of whether the condition $h \geq \varepsilon^{k}$ holds for some integer $k$.
However, the convergence order of ES4 definitely depends on the relation
between $h$ and $\varepsilon$, as shown in \eqref{ES4-error}. That is, ES4 is of order four
when $h \geq \varepsilon^{3}$, while it reduces to third order in the small stepsize regime
$h < \varepsilon^{3}$. This is why ES4 is called a \textit{pseudo-fourth-order} method
in \cite{Niu2026}.

Furthermore, due to the presence of the term $\mathcal{O}(\varepsilon^{4} h^{4})$ in $\text{ED2}_{\mathrm{PN}}$
and ES2, the theory of composition methods (see in Chap. III.5.4 on pp. 92 of \cite{Hairer2006})  implies that this term
cannot be eliminated from the resulting method if $\text{ED2}_{\mathrm{PN}}$ alone
is used to construct a high-order method without its adjoint.
This means that, by following the approach described in \cite{Niu2026},
one can only construct pseudo-high-order methods whose order
reduces to third order in the small stepsize regime where $h < \varepsilon^{k}$
holds for some integer $k$.

\subsection{Novel Hamiltonian splitting and high-order methods}
To achieve genuine fourth-order or even higher-order accuracy regardless of
the relation between $h$ and $\varepsilon$,
we first extend the Hamiltonian in a
special way such that the extended
Hamiltonian is not just a doubled
copy of the original Hamiltonian.
Then, we split the extended Hamiltonian
into three parts, under which high-order explicit
symplectic integrators without order reduction
can be derived.

Following the traditional extension
way established in \cite{Pihajoki2015,Mei2025},
the nonseparable Hamiltonian $H(\bm{p},\bm{q})$
defined in \eqref{perturbatedHamilton} is usually
extended or doubled as:
\begin{equation*}\label{usu-double}
	\widetilde{\Gamma}(\bm{p},\bm{q},\bm{x},\bm{y}) = H_0(\bm{p},\bm{y})
	+ H_0(\bm{x},\bm{q}) +\varepsilon H_1(\bm{p},\bm{y}) +
	\varepsilon H_1(\bm{x},\bm{q}).
\end{equation*}
Here, we extend $H(\bm{p},\bm{q})$ to the doubled
phase space $\mathbb{R}^{4d}$ in a special way as follows:
\begin{equation}\label{double-Hamilton}
	\Gamma(\bm{p},\bm{q},\bm{x},\bm{y}) = H_0(\bm{p},\bm{q})
	+ H_0(\bm{x},\bm{y}) + \varepsilon H_1(\bm{p},\bm{y}) +
	\varepsilon H_1(\bm{x},\bm{q}).
\end{equation}
Let  $F = -\frac{\partial H_0}{\partial \bm{q}}$ and $G = +\frac{\partial H_0}{\partial \bm{p}}$.
The canonical
equations corresponding to the Hamiltonian \eqref{double-Hamilton} read
\begin{equation}\label{eq:doubled_equation}
	\left\{
	\begin{aligned}
		\frac{\mathrm{d}\bm{p}}{\mathrm{d}t} &=
		F(\bm{p}, \bm{q}) + \varepsilon f(\bm{x}, \bm{q}),\\
		\frac{\mathrm{d}\bm{q}}{\mathrm{d}t} &=
		G(\bm{p}, \bm{q}) + \varepsilon g(\bm{p}, \bm{y}),
		\\
		\frac{\mathrm{d}\bm{x}}{\mathrm{d}t} &=
		F(\bm{x}, \bm{y}) + \varepsilon f(\bm{p}, \bm{y}),\\
		\frac{\mathrm{d}\bm{y}}{\mathrm{d}t} &=
		G(\bm{x}, \bm{y}) + \varepsilon g(\bm{x}, \bm{q}).
	\end{aligned}
	\right.
\end{equation}
If the system \eqref{eq:doubled_equation} is subject to the initial condition
\begin{equation*}
	(\bm{p}(0),\bm{q}(0),\bm{x}(0),\bm{y}(0)) = (\bm{p}_0,\bm{q}_0, \bm{p}_0,\bm{q}_0),
\end{equation*}
then the solution of \eqref{eq:doubled_equation}
satisfies $(\bm{p}(t),\bm{q}(t)) = (\bm{x}(t),\bm{y}(t))$, which is just the solution of the original Hamiltonian $H(\bm{p}, \bm{q})$
with the initial values $(\bm{p}_0,\bm{q}_0)$. This establishes the consistency between the
extended Hamiltonian system and the original one.
It is easily noted that $\Gamma(\bm{p},\bm{q},\bm{x},\bm{y})$ is not simply a doubled copy
of $H(\bm{p}, \bm{q})$ and $\Gamma(\bm{p},\bm{q},\bm{x},\bm{y})\neq
\widetilde{\Gamma}(\bm{p},\bm{q},\bm{x},\bm{y})$.

In addition to the previous splitting where $H_A = H_1(\bm{p},\bm{y})$ and
$H_B = H_1(\bm{x},\bm{q})$, we further set $H_C = H_0(\bm{p},\bm{q}) + H_0(\bm{x},\bm{y})$.
This yields a special Hamiltonian splitting for $\Gamma(\bm{p},\bm{q},\bm{x},\bm{y})$:
\begin{equation}\label{special-split}
	\Gamma(\bm{p},\bm{q},\bm{x},\bm{y}) = H_C + \varepsilon H_A + \varepsilon H_B.
\end{equation}
Since $H_C$ is precisely a doubled copy of two independent Kepler flows
for the two-body problem, its complete integrability follows directly
from that of the Kepler problem. Therefore, the splitting \eqref{special-split}
constitutes a separable Hamiltonian splitting, under which each
sub-Hamiltonian is completely integrable and easily solvable.

According to \eqref{subflows}, the phase flows of $\varepsilon H_A$ and $\varepsilon H_B$ are given by
\begin{equation}\label{eq:flows}
	\begin{aligned}
		\exp(\varepsilon t X_{A}) &: (\bm{p}_0, \bm{q}_0,\bm{x}_0, \bm{y}_0)
		\mapsto
		\big(
		\bm{p}_0,\;
		\bm{q}_0 + \varepsilon t g(\bm{p}_0, \bm{y}_0),
		\bm{x}_0 + \varepsilon t f(\bm{p}_0, \bm{y}_0),\;
		\bm{y}_0
		\big),
		\\
		\exp(\varepsilon t X_{B}) &: (\bm{p}_0, \bm{q}_0, \bm{x}_0, \bm{y}_0)
		\mapsto
		\big(
		\bm{p}_0 + \varepsilon t f(\bm{x}_0, \bm{q}_0),\;
		\bm{q}_0 ,
		\bm{x}_0,\;
		\bm{y}_0  + \varepsilon t g(\bm{x}_0, \bm{q}_0)
		\big).
	\end{aligned}
\end{equation}
For $H_C$, the phase flow can be expressed as
\begin{equation}\label{eq:HC_flow}
	\exp(t X_C) : (\bm{p}_0, \bm{q}_0, \bm{x}_0, \bm{y}_0) \mapsto \big(
	\phi_t(\bm{p}_0, \bm{q}_0),\; \phi_t(\bm{x}_0, \bm{y}_0) \big),
\end{equation}
where $X_C = \{\,\cdot\,,H_C\}=\frac{\partial H_0}{\partial \bm{p}}
\frac{\partial }{\partial \bm{q}}-\frac{\partial H_0}{\partial \bm{q}}\frac{\partial }{\partial \bm{p}}
+\frac{\partial H_0}{\partial \bm{x}}\frac{\partial }{\partial \bm{y}}
-\frac{\partial H_0}{\partial \bm{y}}\frac{\partial }{\partial \bm{x}}$
and $\phi_t$ is the phase flow of the Hamiltonian $H_0(\bm{p},\bm{q})$.

Based on this splitting and following the symmetric composition with
the phase flows $\exp(\varepsilon h X_{A})$, $\exp(\frac{\varepsilon h}{2} X_{B})$, and $\exp(\frac{h}{2} X_{C})$ in
St\"{o}rmer--Verlet scheme, we obtain a second-order explicit
doubled phase space symplectic integrator:
\begin{eqnarray}
	\mathrm{DS2}^{*}(h) =
	&&\exp(\tfrac{h}{2} X_{C}) \circ \exp(\tfrac{\varepsilon h}{2} X_{B})
	\circ \exp(\varepsilon h X_{A})
	\nonumber
	\\
	&& \circ \exp(\tfrac{\varepsilon h}{2} X_{B}) \circ \exp(\tfrac{h}{2} X_{C}).
	\label{eq:part_DS2}
\end{eqnarray}
Then, adding the projection operator \eqref{eq:ED2_projection_row} to $\mathrm{DS2}^*$ finally yields
the new explicit method $\mathrm{DS2}$ as follows:
\begin{eqnarray}
	\mathrm{DS2}(h) =
	&&\mathcal{P} \circ \exp(\tfrac{h}{2} X_{C}) \circ \exp(\tfrac{\varepsilon h}{2} X_{B}) \circ \exp(\varepsilon h X_{A}) \circ \exp(\tfrac{\varepsilon h}{2} X_{B}) \circ \exp(\tfrac{h}{2} X_{C}).
	\label{eq:full_DS2_definition}
\end{eqnarray}
Similar to \eqref{eq:ES4}, applying Yoshida's triple symmetric
composition to $\mathrm{DS2}^*$, merging neighboring
phase flows, and adding the projection operator $\mathcal{P}$
finally yields a fourth-order explicit method:
\begin{eqnarray}
	\text{DS4}(h)=
	&&\mathcal{P}\circ\exp(\tfrac{\lambda h}{2} X_{C}) \circ
	\exp(\tfrac{\lambda\varepsilon h}{2} X_{B})
	\circ \exp(\lambda\varepsilon h X_{A})\circ
	\exp(\tfrac{\lambda\varepsilon h}{2} X_{B})
	\circ\exp(\tfrac{(1-\lambda)h}{2} X_{C})\nonumber
	\\
	&&\circ
	\exp(\tfrac{(1-2\lambda)\varepsilon h}{2} X_{B})
	 \circ\exp((1-2\lambda)\varepsilon h X_{A}) \circ
	\exp(\tfrac{(1-2\lambda)\varepsilon h}{2} X_{B})
    \circ\exp(\tfrac{(1-\lambda)h}{2} X_{C})\nonumber
	\\
	&&\circ
    \exp(\tfrac{\lambda\varepsilon h}{2} X_{B})
	\circ \exp(\lambda\varepsilon h X_{A})
	\circ \exp(\tfrac{\lambda\varepsilon h}{2} X_{B})
	\circ \exp(\tfrac{\lambda h}{2} X_{C}).
	\label{eq:DS4}
\end{eqnarray}
Higher-order explicit methods can be obtained in a similar
way by using operator-splitting theory based on the BCH formula.

We denote the numerical solution at the $n$th step by $(\bm{p}_n,\bm{q}_n)$ such that
$(\bm{p}_n,\bm{q}_n)\approx (\bm{p}(t_n),\bm{q}(t_n))$ with $t_n=nh$. Then the detailed numerical scheme
of $\mathrm{DS2}$ for time-stepping from $(\bm{p}_n,\bm{q}_n)$ to $(\bm{p}_{n+1},\bm{q}_{n+1})$ is as follows:

\begin{equation}\label{ext-leapfrog}
	\left\{
	\begin{aligned}
		&(\bm{p}^{(1)},\bm{q}^{(1)})
		= \phi_{h/2}(\bm{p}_n, \bm{q}_n),
		\\
		&\bm{p}^{(2)}
		= \bm{p}^{(1)} + \tfrac{\varepsilon h}{2} f(\bm{p}^{(1)},\bm{q}^{(1)}),
		\\
		&\bm{y}^{(2)}
		= \bm{q}^{(1)} + \tfrac{\varepsilon h}{2} g(\bm{p}^{(1)},\bm{q}^{(1)}),
		\\
		&\bm{x}^{(2)}
		= \bm{p}^{(1)} + \varepsilon h f(\bm{p}^{(2)},\bm{y}^{(2)}),
		\\
		&\bm{q}^{(2)}
		= \bm{q}^{(1)} + \varepsilon h g(\bm{p}^{(2)},\bm{y}^{(2)}),
		\\
		&\bm{p}^{(3)}
		= \bm{p}^{(2)} + \tfrac{\varepsilon h}{2} f(\bm{x}^{(2)},\bm{q}^{(2)}),
		\\
		&\bm{y}^{(3)}
		= \bm{y}^{(2)} + \tfrac{\varepsilon h}{2} g(\bm{x}^{(2)},\bm{q}^{(2)}),
		\\
		&(\bm{x}^{(3)}, \bm{q}^{(3)}) = \phi_{h/2}(\bm{x}^{(2)}, \bm{q}^{(2)}),
		\\
		&(\bm{p}^{(4)}, \bm{y}^{(4)}) = \phi_{h/2}(\bm{p}^{(3)}, \bm{y}^{(3)}),
		\\
		&(\bm{p}_{n+1}, \bm{q}_{n+1},\bm{p}_{n+1},\bm{q}_{n+1})
		= \mathcal{P} \bigl(\bm{p}^{(4)},\bm{q}^{(3)}, \bm{x}^{(3)},\bm{y}^{(4)}\bigr).
	\end{aligned}
	\right.
\end{equation}
The numerical scheme of DS4 can be obtained similarly and is omitted here.

We next consider the local truncation error of the new method DS2.
Let $(\bm{p}_0, \bm{q}_0,\bm{p}_0, \bm{q}_0)$ be the input for
the doubled Hamiltonian $\Gamma(\bm{p},\bm{q},\bm{x},\bm{y})$,
and let $(\tilde{\bm{p}}_1,\tilde{\bm{q}}_1,\tilde{\bm{x}}_1,\tilde{\bm{y}}_1)$
be the numerical solution obtained after advancing one step with $\mathrm{DS2}^{*}$.
Due to the special Hamiltonian splitting defined in \eqref{special-split}
for $\Gamma(\bm{p},\bm{q},\bm{x},\bm{y})$ and the second-order accuracy of the
St\"{o}rmer--Verlet scheme $\mathrm{DS2}^{*}$, we have
\begin{equation}\label{eq:local-err}
	(\tilde{\bm{p}}_1,\tilde{\bm{q}}_1,\tilde{\bm{x}}_1,\tilde{\bm{y}}_1) -
	(\bm{p}(h), \bm{q}(h),\bm{p}(h), \bm{q}(h)) = \mathcal{O}(\varepsilon h^3),
\end{equation}
where $(\bm{p}(h), \bm{q}(h),\bm{p}(h), \bm{q}(h))$ is the exact solution of \eqref{eq:doubled_equation}
at time $t=h$ with the initial values $(\bm{p}_0, \bm{q}_0,\bm{p}_0, \bm{q}_0)$.

It immediately follows from \eqref{eq:local-err} that
\begin{align*}
	(\tilde{\bm{p}}_1,\tilde{\bm{q}}_1) - (\bm{p}(h), \bm{q}(h))
	= \mathcal{O}(\varepsilon h^3),
	(\tilde{\bm{x}}_1,\tilde{\bm{y}}_1) - (\bm{p}(h), \bm{q}(h))
	= \mathcal{O}(\varepsilon h^3).
\end{align*}
Let $(\bm{p}_1,\bm{q}_1)$ be the numerical solution obtained by $\mathrm{DS2}$ with stepsize $h$, i.e.,
\begin{equation*}
	\mathrm{DS2}(h): (\bm{p}_0, \bm{q}_0) \mapsto (\bm{p}_1, \bm{q}_1).
\end{equation*}
Then, we have
\begin{equation*}
   (\bm{p}_1, \bm{q}_1,\bm{p}_1, \bm{q}_1) = \mathcal{P}
   (\tilde{\bm{p}}_1,\tilde{\bm{q}}_1,\tilde{\bm{x}}_1,\tilde{\bm{y}}_1).
\end{equation*}
According to the definition of the projection operator $\mathcal{P}$ given
in \eqref{eq:ED2_projection_row}, we further obtain
\begin{equation*}
	(\bm{p}_1, \bm{q}_1) - (\bm{p}(h), \bm{q}(h))
	= \mathcal{O}(\varepsilon h^3),
\end{equation*}
which is the local truncation error of $\mathrm{DS2}$. Likewise, we can derive
that the local truncation error of $\mathrm{DS4}$ is $\mathcal{O}(\varepsilon h^5)$,
which has fourth-order accuracy in all stepsize regime
and is independent of the relation between $h$ and $\varepsilon$.    	

\section{Numerical experiments}
In this section, we carry out numerical experiments
to test the performance of the proposed explicit methods.
We select the conservative 2PN Hamiltonian system of spinning
compact binaries as the test model for convenience. For the
dissipative PN Hamiltonian including the gravitational radiation
effect, the applicability still holds, since it can be extended
to a conservative system by adding an auxiliary pair of
canonical variables. As noted in \cite{Niu2026}, the proposed
methods are also applicable to the PN $n$-body problem,
such as the Solar System with PN corrections, by applying
the $H_0+\varepsilon H_1$ splitting technique proposed in
\cite{Huang2014} for the Hamiltonian under consideration.

\subsection{Dynamical model of spinning compact binaries}\label{sec:formula}
The 2PN Hamiltonian is formulated in the center-of-mass frame using
Arnowitt--Deser--Misner (ADM) coordinates.
Suppose the two compact bodies have masses $m_1$ and $m_2$
($m_1 \leq m_2$). Let $\bm{Q}$
be the relative position vector from body 2 to body 1,  $\bm{P}$
be the conjugate momentum of body 1 relative to the center of
mass, and $\bm{S}_i$ be the spin of body $i$ for $i=1,2$.
To conveniently formulate the Hamiltonian, we further introduce
the total mass $M=m_1 + m_2$, the mass ratio $\beta = m_1/m_2$,
the reduced mass $\mu = m_1m_2/M$,
the parameter $\eta = \mu/M = \beta/(1+\beta)^2$, the separation distance
$r = |\bm{Q}| $, and the unit radial vector $\bm{N} = \bm{Q}/r$.

Under the normalization $G=1$ and $M=1$, the PN Hamiltonian reads:
\begin{equation}\label{eq:H_full}
	H = H_{\mathrm{N}} + \frac{1}{c^2} H_{\mathrm{1PN}}
	+ \frac{1}{c^3} H_{\mathrm{1.5PN}}^{\mathrm{SO}}
	+ \frac{1}{c^4} \left( H_{\mathrm{2PN}}
	+ H_{\mathrm{2PN}}^{\mathrm{SS}} \right),
\end{equation}
where the individual contributions are given as
follows~\cite{Buonanno2006,Hartl2005,Levin2006,Nagar2011}:
\begin{equation*}
	\begin{aligned}
		&H_{\mathrm{N}}=\frac{\bm{P}^2}{2}-\frac{1}{r},
		\\
		&H_{\mathrm{1PN}}=\frac{1}{8}(3\eta-1)\bm{P}^4-\frac{1}{2}\big[(3+\eta)\bm{P}^2
		+\eta(\bm{N}\cdot\bm{P})^2\big]\frac{1}{r} +\frac{1}{2r^{2}},
		\\
		&H_{\mathrm{2PN}}=
		\frac{1}{16}(1-5\eta+5\eta^2)\bm{P}^6
		+\frac{1}{8}\big[(5-20\eta-3\eta^2)\bm{P}^4-2\eta^2{(\bm{N}\cdot\bm{P})^2}\bm{P}^2 -3\eta^2{(\bm{N}\cdot\bm{P})^4}\big]\frac{1}{r}
		\\
		&\qquad\qquad+\frac{1}{2}\big[(5+8\eta)\bm{P}^2
		+3\eta (\bm{N}\cdot\bm{P})^2\big]\frac{1}{r^2}-\frac{1}{4}(1+3\eta)\frac{1}{r^3},
		\\
		&H_{\mathrm{1.5PN}}^{\mathrm{SO}} = \frac{1}{r^{3}} \Bigl((2+\tfrac{3}{2\beta})\bm{S}_1
		+ (2+ \tfrac{3\beta}{2})\bm{S}_2\Bigr)\cdot\bm{L},
		\\
		&H_{\mathrm{2PN}}^{\mathrm{SS}}= \frac{1}{2r^3}
		\Bigl[3\Big(\big((1+\tfrac{1}{\beta})\bm{S}_1+(1+\beta)\bm{S}_2)\big)\cdot\bm{N}\Big)^2-\big((1+\tfrac{1}{\beta})\bm{S}_1+(1+\beta)\bm{S}_2)\big) ^2\Bigr],
	\end{aligned}
\end{equation*}
where $ H_\mathrm{N} $ is the classical Newtonian term, $ H_{\mathrm{1PN}} $ and $ H_{\mathrm{2PN}} $
are the orbital contributions at 1PN and 2PN orders, respectively,
$ H_{\mathrm{1.5PN}}^{\mathrm{SO}} $ denotes the spin-orbit coupling at 1.5PN order, and
$ H_{\mathrm{2PN}}^{\mathrm{SS}}$ represents the spin-spin coupling at 2PN order.

As stated in \cite{HuangLi2018,Huang2019,Niu2026}, the velocity of light $c$
should be adjusted to be consistent with the normalization $G=1$ and $M=1$.
A typical example is the Sun-Earth system, where $c$ should be rescaled to
$c\approx 1.0067\times10^4$ using AU as the unit of length. To reflect different
scales of the PN effect, we may readjust the value of $c$. We also note that
$t$, $\bm{Q}$, $\bm{P}$, and $\bm{S}_i$ are not physical but rescaled
variables, obtained as $t \rightarrow t/GM$, $\bm{Q} \rightarrow \bm{Q}/GM$,
$\bm{P} \rightarrow \bm{P}/\mu$, $\bm{S} \rightarrow \bm{S}/M^2$
(see, e.g., \cite{Hartl2005,Levin2006}). The spin magnitudes are usually
measured by $\Lambda_i = \chi_{i} m_{i}^{2} / M^{2}$, where $\chi_i\in [0, 1]$.

However, the symplectic integrator cannot be directly applied
to the Hamiltonian \eqref{eq:H_full} since the spins $\bm{S}_i$
are not canonical variables. Here, we adopt the canonical
conjugate spin variables $\bm{\theta}=(\theta_1,\theta_2)$ and
$\bm{\xi}=(\xi_1,\xi_2)$ introduced by Wu \& Xie \cite{Wu2010} as follows:
\begin{equation}\label{spin-new-v}
	\bm{S}_i=\left(
	\begin{aligned}
		&\sqrt{\Lambda_i^2-\xi_i^2}\cos(\theta_{i})
		\\
		&\sqrt{\Lambda_i^2-\xi_i^2}\sin(\theta_{i})
		\\
		&\qquad \xi_{i}
	\end{aligned}
	\right),\quad i=1,2,
\end{equation}
where the spin magnitudes $\Lambda_i =|\bm{S}_i| $ are conserved during
the evolution of the system. Then, with these new conjugate variables
$(\bm{P},\bm{Q},\bm{\xi},\bm{\theta})$, the evolution equations
can be written canonically as
\begin{equation}\label{eq:canonical_eom}
	\begin{aligned}
		\frac{\mathrm{d}\bm{P}}{\mathrm{d}t} &= -\frac{\partial H}{\partial \bm{Q}},
		\quad \frac{\mathrm{d}\bm{Q}}{\mathrm{d}t} &= +\frac{\partial H}{\partial \bm{P}}, \\
		\frac{\mathrm{d}\bm{\xi}}{\mathrm{d}t} &= -\frac{\partial H}{\partial \bm{\theta}},
		\quad \frac{\mathrm{d}\bm{\theta}}{\mathrm{d}t} &= +\frac{\partial H}{\partial \bm{\xi}}.
	\end{aligned}
\end{equation}
In addition to the total energy
$H(\bm{Q},\bm{P},\bm{\theta},\bm{\xi})$,
there are other first integrals for the system, namely the total
angular momentum vector
\begin{equation*}
	\bm{J}=\bm{L}+\bm{S}_1+\bm{S}_2,
\end{equation*}
where $\bm{L}=\bm{Q}\times\bm{P}$
is the orbital angular momentum vector.

To apply the proposed explicit symplectic methods
DS2 and DS4, we need to split the 2PN Hamiltonian
$H(\bm{P},\bm{Q},\bm{\xi},\bm{\theta})$ into two parts as in \eqref{perturbatedHamilton}.
Here, we employ the natural splitting such that $H_0 = H_{\mathrm{N}}$ and
\begin{equation}
	H_{1}=H_{\mathrm{1PN}} + \varepsilon H_{\mathrm{2PN}}+\sqrt{\varepsilon}
	H_{\mathrm{1.5PN}}^{\mathrm{SO}}+ \varepsilon H_{\mathrm{2PN}}^{\mathrm{SS}},
\end{equation}
where $\varepsilon=1/c^2$. To assess the feasibility and efficacy of our
newly proposed integrators, we evaluate the global Hamiltonian
error (GHE), the global angular momentum error (GJ),
and the global error (GE) in the variables $(\bm{Q}, \bm{P}, \bm{\theta}, \bm{\xi})$, with $|\bm{J}| = \sqrt{\bm{J}^2}$.

\subsection{Numerical results}\label{sec:numerical}
The following integrators are selected for comparison:
\begin{itemize}
	\item DS2/DS4: the novel second- and fourth-order explicit
	symplectic methods given by~\eqref{eq:full_DS2_definition} and~\eqref{eq:DS4}
	with $(\lambda_0,\mu_0)=(\frac{1}{e},\frac{1}{\pi})$;
	\item ES2/ES4: the second- and fourth-order explicit symplectic
	methods given in \cite{Niu2026} with
	$(\lambda_0,\mu_0)=(\frac{1}{e},\frac{1}{\pi})$;
	\item IM2/IM4: the second- and fourth-order semiexplicit
	symmetric projection symplectic methods proposed in~\cite{Jayawardana2023};
	\item IRK2/IRK4: the second- and fourth-order symplectic Gauss
	collocation Runge--Kutta methods.
\end{itemize}
All methods except IRK2/IRK4 evolve in the doubled phase space.
Among them, ES2/ES4 and DS2/DS4 are fully explicit schemes that
inherently incorporate the post-Newtonian expansion parameter
$\varepsilon$, whereas IM2/IM4 and IRK2/IRK4 are implicit and do not take
the small parameter $\varepsilon$ into account. That is, the methods ES2/ES4
and DS2/DS4 are expected to have higher accuracy than their
same-order counterparts IM2/IM4 and IRK2/IRK4. In practice, the
fixed-point iteration is employed for IRK2/IRK4, while a simplified
Newton scheme is used for IM2/IM4 \cite{Jayawardana2023},
both with a uniform iteration tolerance of $10^{-12}$.
To measure the global error, we use the numerical solution obtained by
the 8th-order symplectic Gauss collocation RK method with a
tiny time stepsize as the reference solution.

We consider an orbit with initial conditions $\bm{Q}(0) = (25.34,\, 0,\, 0)$,
$\bm{P}(0) = (0,\, 0.18,\, 0)$, $\bm{\theta}(0) = (1.2490,\, 0.6202)$, and
$\bm{\xi}(0) = (0.0445,\, 0.6104)$, as well as parameters
$(\Lambda_1,\, \Lambda_2) = (0.0479,\, 0.6104)$ and $\beta = 0.28$.
We take two values of $c$: $c_1 = 10^{1/2}$ and $c_2 = 10^{1/10}$,
which correspond to $\varepsilon_1 = 0.1$ and $\varepsilon_2 \approx 0.63$, respectively.
Here, we adopt the same setting $(\lambda_0,\mu_0) = (\frac{1}{e},\frac{1}{\pi})$ as
in \cite{Niu2026}, although their values are only
required to lie in the interval $(0,1)$.

\begin{figure*}[htb]
	\centering{
		\includegraphics[scale=0.4]{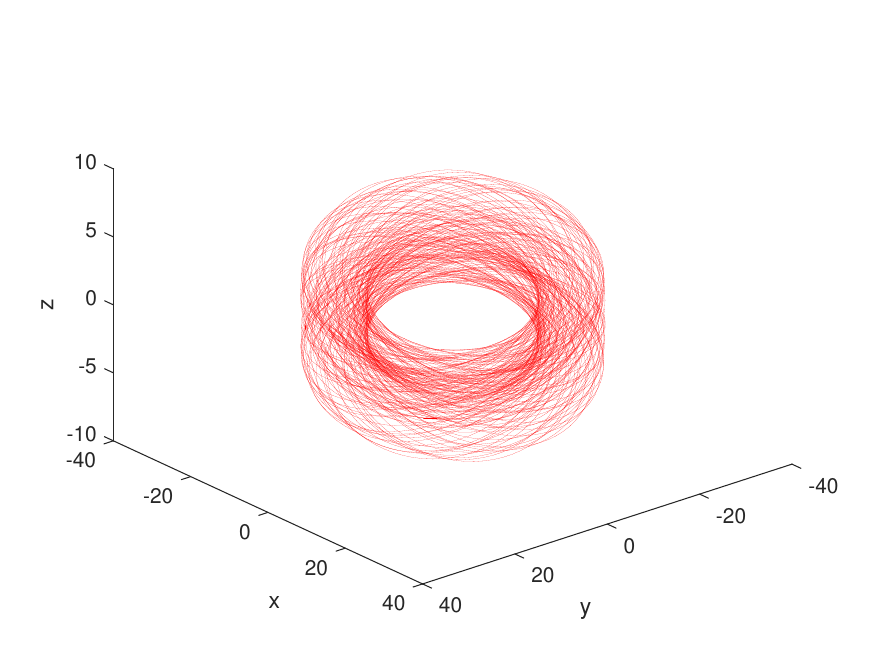}
		\includegraphics[scale=0.4]{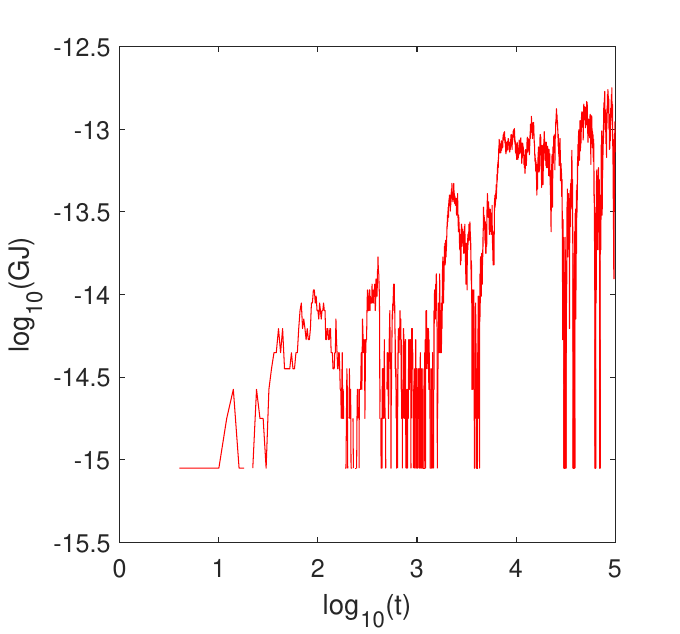}
		\includegraphics[scale=0.4]{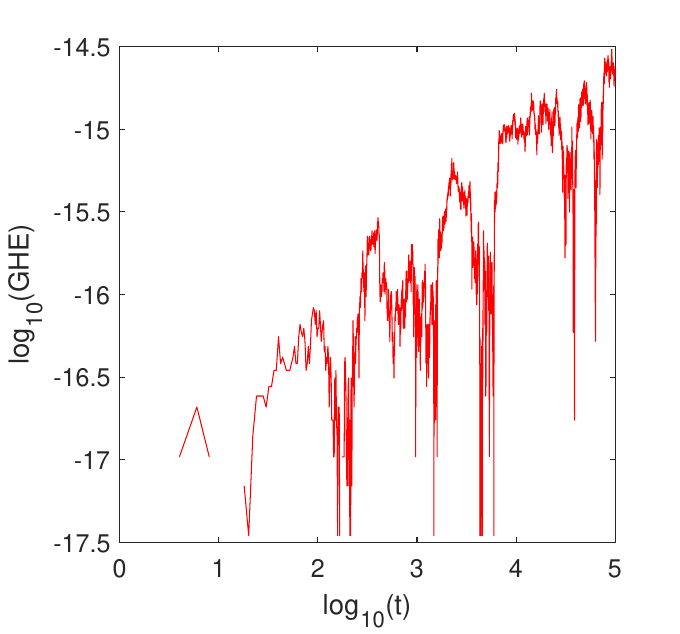}
	}
	\caption{\label{fig1}
		Orbital trajectory (left), global angular momentum error (middle),
		and global Hamiltonian error (right) for the reference solution.}
\end{figure*}

Fig.~\ref{fig1} displays, from left to right, the regular reference
trajectory, the global angular momentum error (GJ), and the global
Hamiltonian error (GHE) over an integration time of $T = 10^5$
with a step size of $h = 1$. As shown in this figure,
the reference solution achieves high accuracy.

\begin{figure*}[htb]
	\centering
	\includegraphics[scale=0.4]{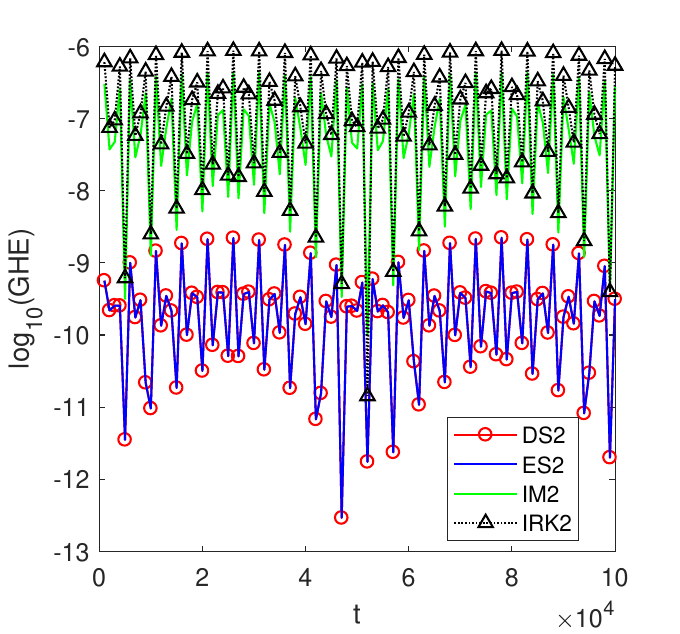}
	\includegraphics[scale=0.4]{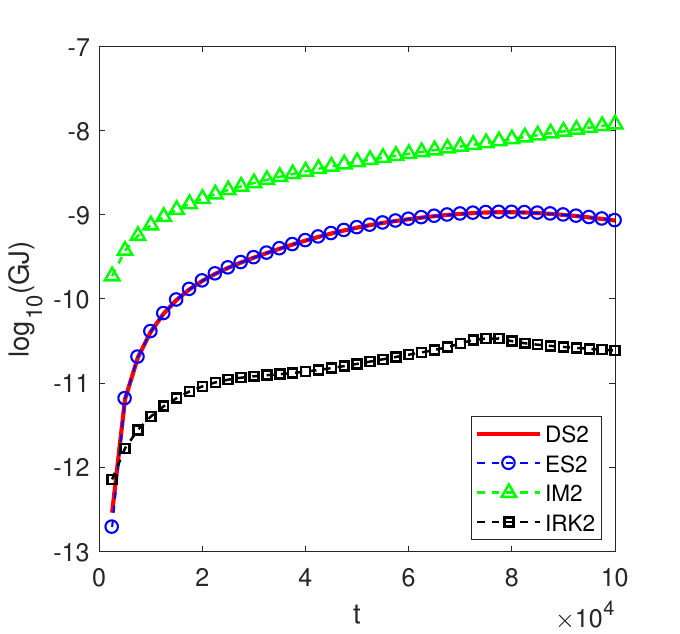}
	\includegraphics[scale=0.4]{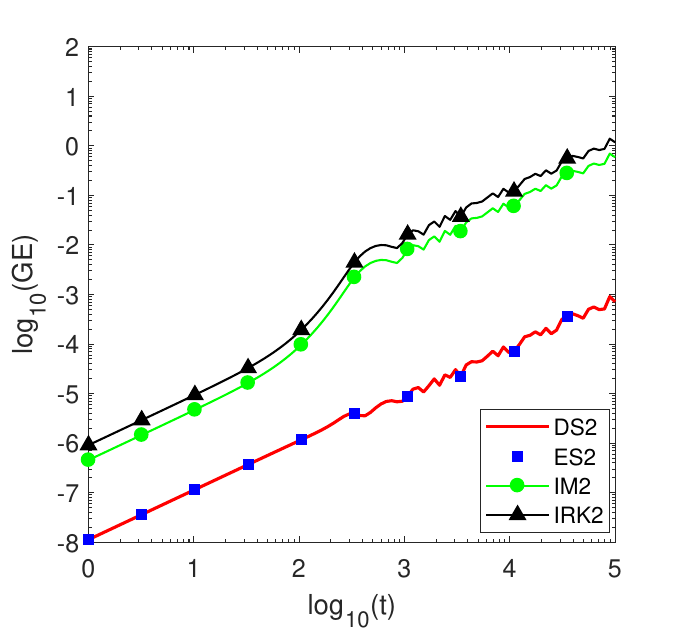}
	\includegraphics[scale=0.4]{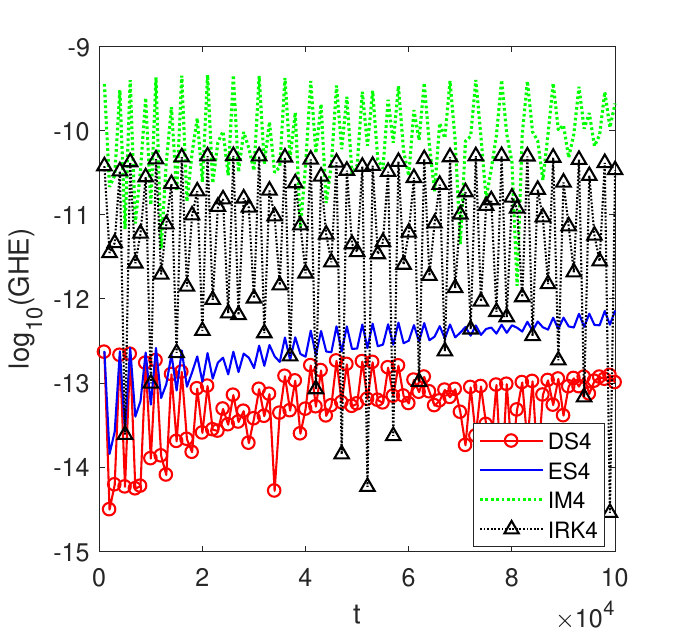}
	\includegraphics[scale=0.4]{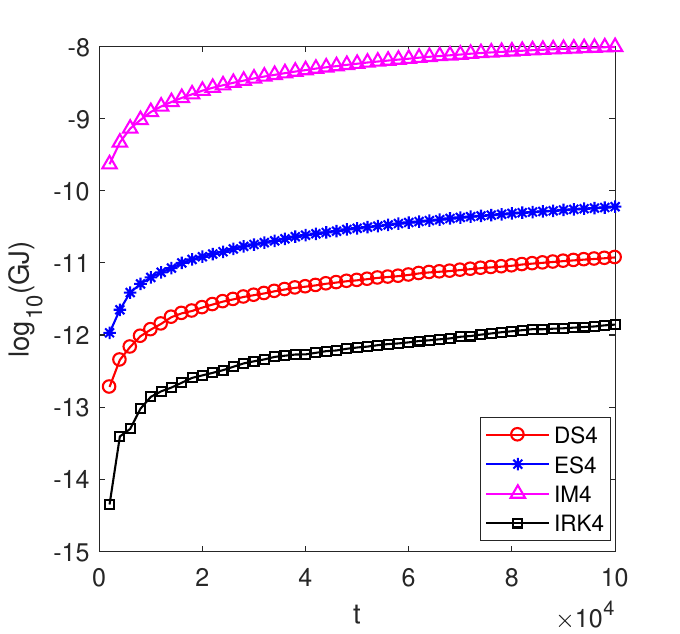}
	\includegraphics[scale=0.4]{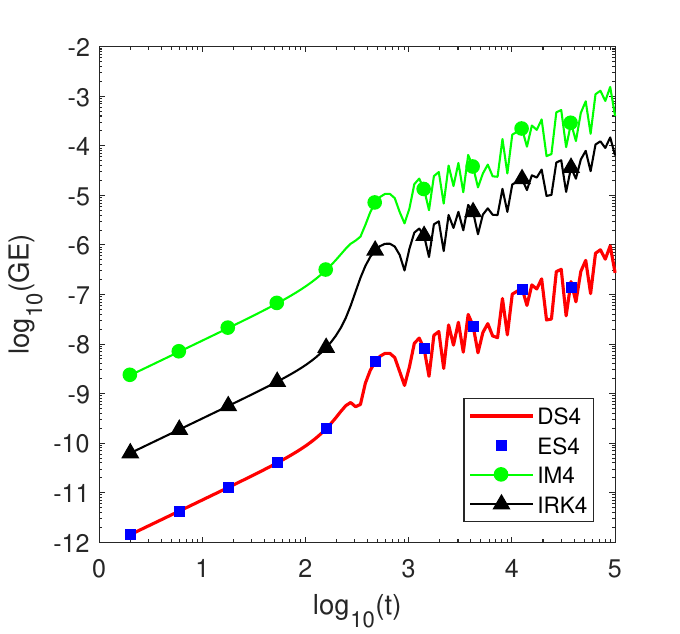}
	\caption{\label{fig2}
		Global Hamiltonian error (left), global angular momentum error (middle),
		and global error (right) with $h=1$ for second-order methods
		and with $h=2$ for fourth-order methods.}
\end{figure*}
\medskip

Fig.~\ref{fig2} illustrates the global Hamiltonian error, global
angular momentum error, and global error for second-order
methods (with a stepsize $h=1$) and fourth-order methods
(with a stepsize $h=2$) over an
integration interval of $T=10^5$ with parameter $\varepsilon =0.1$.
The results demonstrate that all tested symplectic integrators exhibit
excellent conservation properties: both GHE and GJ remain bounded without
long-term drift. Consistent with the theoretical expectations for symplectic methods,
the global error for all methods displays a linear growth over time.
Furthermore, a comparative analysis reveals that while the specially designed
symplectic methods DS2/DS4 achieve energy and state accuracy comparable
to ES2/ES4 and superior to IM2/IM4 and IRK2/IRK4, the IRK methods attain higher
accuracy in angular momentum conservation, because symplectic RK methods preserve
the quadratic invariant and the quadratic angular momentum $\bm{L}=\bm{Q}\times\bm{P}$
dominates the total angular momentum $\bm{J}$.

\begin{figure*}[htb]
	\centering
	\includegraphics[scale=0.60]{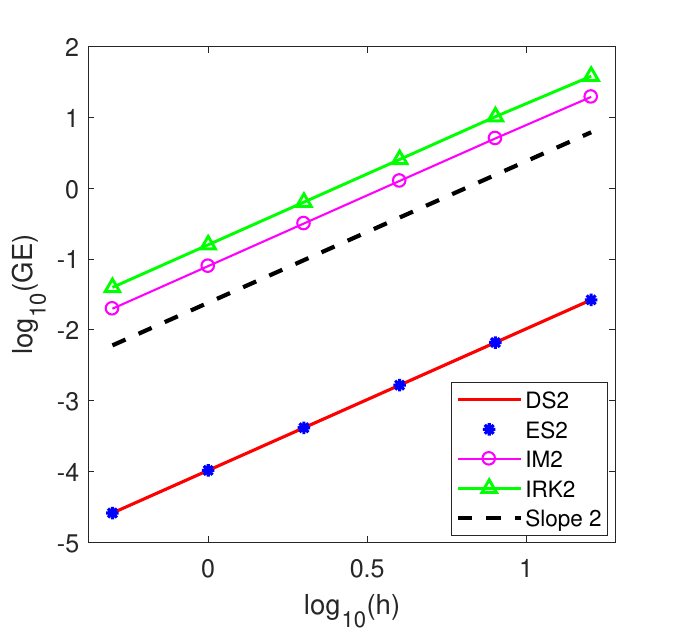}
	\includegraphics[scale=0.60]{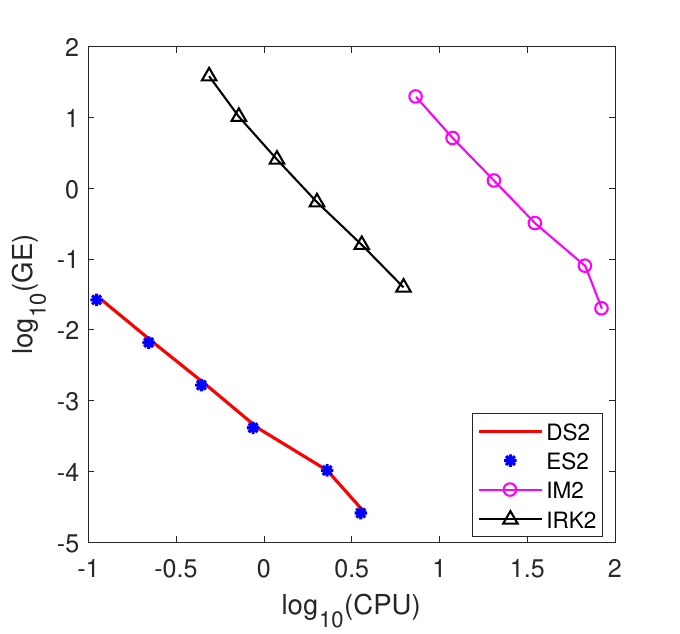}
	\includegraphics[scale=0.60]{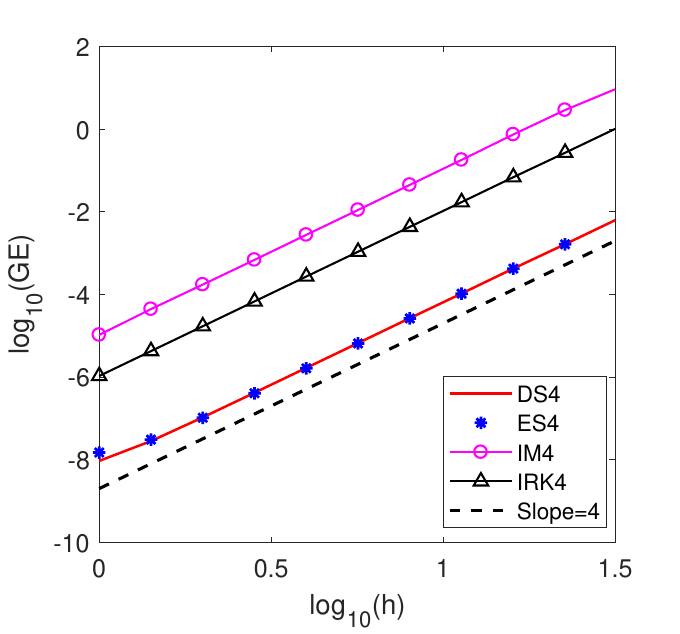}
	\includegraphics[scale=0.60]{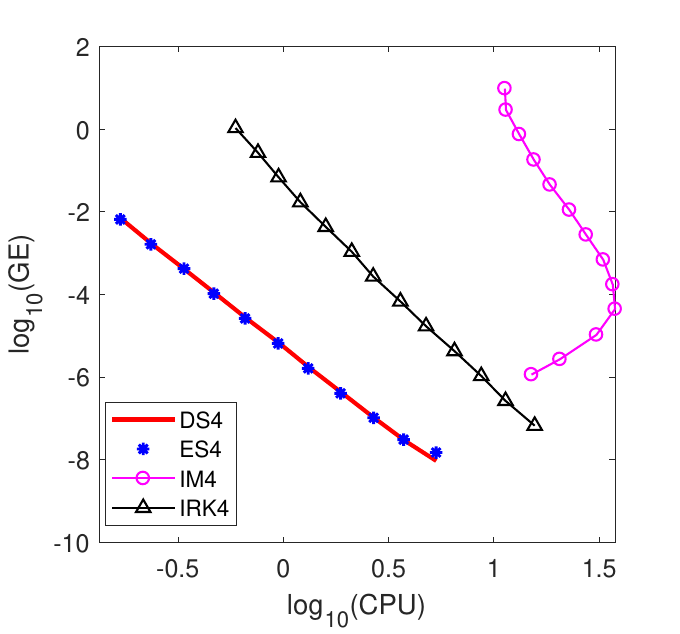}
	\caption{\label{fig3}
		Convergence orders (left) and efficiency curves (right)
		with $\varepsilon=0.1$.}
\end{figure*}

We further test the convergence orders and computational efficiency
of all the symplectic integrators in Fig.~\ref{fig3}, where the parameter
$\varepsilon$ is set to $0.1$. It is observed from this figure that
DS2, ES2, IM2, and IRK2 are of order 2, while DS4, ES4,
IM4, and IRK4 are of order 4. This is consistent with the theoretical
orders of these methods. Notably, the pseudo-fourth-order
method ES4 demonstrates fourth-order behavior in this case.
In terms of performance, the explicit symplectic methods DS2/ES2
and DS4/ES4 consistently yield the smallest global errors and
the highest computational efficiency among methods of the same order,
highlighting the inherent advantages of explicit
formulations for this class of problems.

\begin{figure*}[htb]
	\centering
	\includegraphics[scale=0.6]{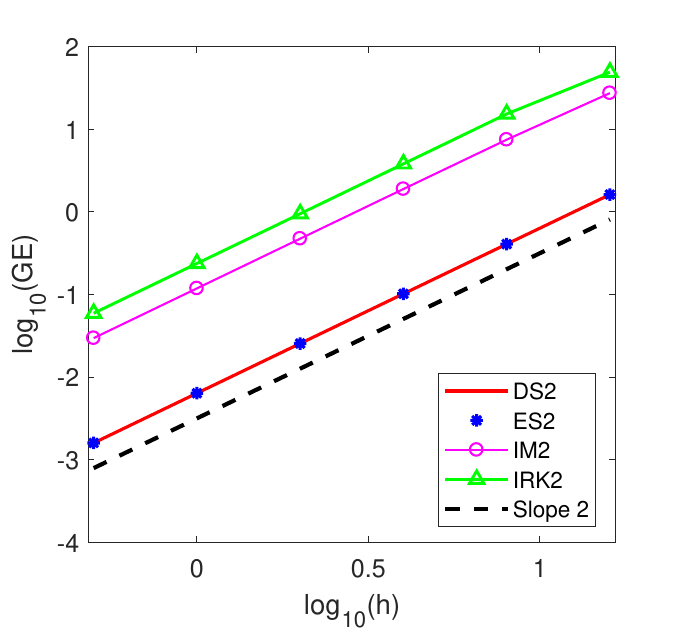}
	\includegraphics[scale=0.6]{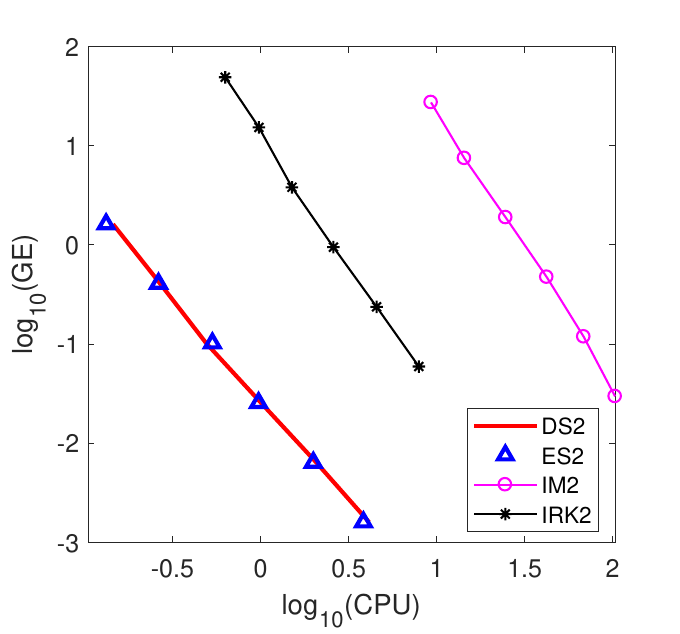}
	\includegraphics[scale=0.6]{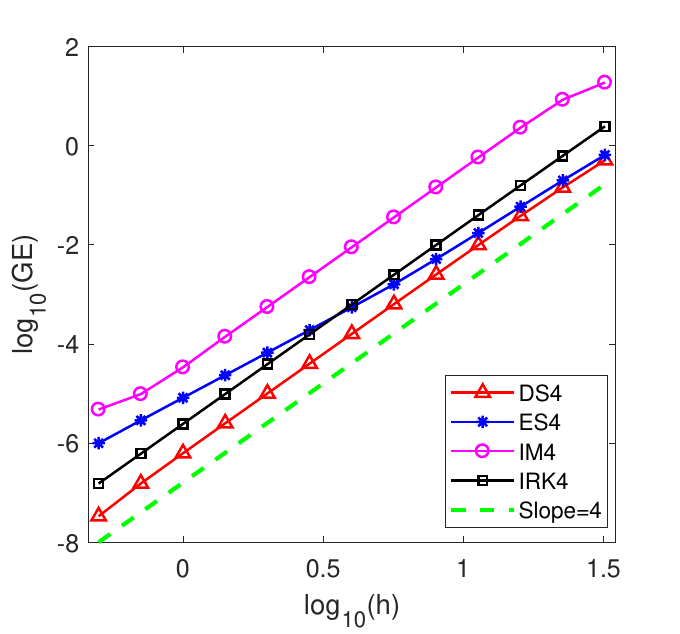}
	\includegraphics[scale=0.6]{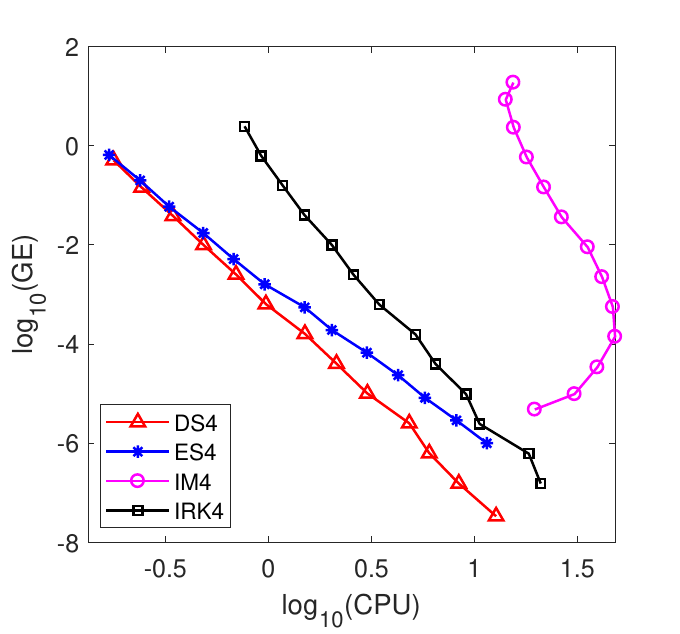}
	\caption{\label{fig4}
		Convergence orders (left) and efficiency curves (right)
		with $\varepsilon\approx0.63$.}
\end{figure*}

To rigorously validate that the newly proposed method
DS4 is genuinely fourth-order while ES4 reduces to
third-order in some cases, we conduct a numerical
experiment with $\varepsilon\approx0.63$ in Fig.~\ref{fig4}, the stepsizes remaining
the same as those used in Fig.~\ref{fig3}. Compared to
Fig.~\ref{fig3}, the most notable point is that the
pseudo-fourth-order method ES4 has an overall
order of $3.12$, which clearly indicates order reduction.
In contrast, the proposed method DS4 robustly
maintains its fourth-order convergence. In this case,
the efficiency of ES4 is also reduced, and the new
method DS4 achieves the highest efficiency among
all fourth-order methods. This confirms the capability
of our new methodology to construct genuine high-order
explicit symplectic integrators that, by considering a
special Hamiltonian splitting, are unaffected by the
stepsize and the small parameter and achieve higher
accuracy than generally designed symplectic methods.

\section{Conclusion}\label{sec:conclusion}

We have developed a novel class of explicit symplectic integrators
for nonseparable post-Newtonian Hamiltonian systems. The methodology
consists of embedding the dynamics into a doubled phase space, numerically
solving the extended system with a special Hamiltonian splitting, and projecting the
numerical solutions from the doubled phase space onto the special submanifold.
Since the doubled Hamiltonian is decomposed in a particular way, the small
parameter $\varepsilon$ is retained in the truncation error, making
the new explicit symplectic integrators achieve higher accuracy than
generally designed symplectic methods of the same order. In particular,
the convergence order of the new methods remains unchanged and independent
of the relation between $h$ and $\varepsilon$. However, the existing explicit method ES4
proposed in \cite{Niu2026} suffers from order reduction for small $h$
(with respect to $\varepsilon$). Numerical experiments on 2PN spinning binaries demonstrate
that our fourth-order method offers superior long-term conservation
properties and computational efficiency compared to traditional implicit schemes
and the explicit method in \cite{Niu2026}.
While this study focuses on conservative 2PN dynamics, the underlying
methodology is readily adaptable to dissipative systems and a broader class
of nonseparable Hamiltonian problems consisting of a dominant
integrable part and a perturbed nonseparable part.
	
\acknowledgments
This work was supported in part by the National Natural Science Foundation
of China (Grant No.~12163003) and  the Yunnan Fundamental Research Projects
(Grant No.~202401CF070033).
	
	

\end{document}